\documentclass[letterpaper,10pt,conference]{ieeeconf}

\IEEEoverridecommandlockouts
\usepackage{amssymb}
\usepackage{graphicx}
\usepackage[cmex10]{amsmath}
\usepackage{array}
\usepackage{mdwtab}
\usepackage{graphics}
\usepackage{epsfig}
\usepackage{times}
\usepackage{amsmath}
\usepackage{amssymb}
\usepackage{url}
\usepackage{lscape}
\usepackage{color}
\usepackage{epsfig}
\usepackage{cite}
\usepackage{csquotes}
\usepackage{balance}
\usepackage{dblfloatfix}
\usepackage{hyperref}
\usepackage{mathtools}

\title{\LARGE \bf Design and Implementation of a Controller for Hexaglide CNC Machine}

\author{Kaveh Safavigerdini\textsuperscript{1}, Aria Alasty\textsuperscript{2}, and Mohammad Reza Movahhedy\textsuperscript{2}
\thanks{
\textsuperscript{1}Kaveh Safavigerdini is with the Department of Electrical and Computer Engineering, University of Missouri, Columbia, MO, USA 65211. \newline
\textsuperscript{2}Aria Alasty and Mohammad Reza Movahhedy are with the Department of Mechanical Engineering, Sharif University of Technology, Tehran, Iran.
}
}

\begin{document}
\renewcommand\thesection{\Roman{section}}
\renewcommand\thesubsection{\thesection.\Roman{subsection}}

\maketitle
\thispagestyle{empty}
\pagestyle{empty}

\maketitle

\begin{abstract}
The objective of this research is to develop an advanced controller for CNC machines equipped with Hexaglide parallel mechanisms. While traditional CNC machines employ mechanisms with perpendicular axes, accuracy may be compromised when processing high force loads. Parallel mechanisms, such as Hexaglide, have gained attention in the CNC industry due to their ability to maintain precision at high speeds and power. However, the absence of a sophisticated controller for converting G-Code to motion paths has been a significant drawback for CNC machines with parallel mechanisms. This research proposes a solution by introducing a controller that can translate G-Code generated by code software into appropriate motion commands for the Hexaglide mechanism.\\
\end{abstract}

%---------------------------------------------------------------------------------------------------------
%---------------------------------------------------------------------------------------------------------
\section{Introduction}\label{sec:introduction}
In the 20th century, the invention of numerical control machines marked a significant advancement in production. As computers evolved, CNC machines were introduced to the market and rapidly gained popularity in the manufacturing and production industry. The development and growth of CNC machines had a tremendous impact on automation in machining processes, allowing for increased flexibility and precision. 
% Both traditional and modern controllers are commonly utilized in CNC machines, with examples including the conventional proportional-integral-derivative (PID) controller and newer variations like fractional PID controllers \cite{doi:10.1177/0959651819849284, yaghooti2022stabilizing}.
In modern machinery, the need for mechanisms with improved dynamics is becoming increasingly necessary due to the rise in machine speed and the need for proper load-bearing capabilities. One effective solution for this is the use of parallel mechanisms in CNC machines\cite{lu2007kinematics}. In recent years, parallel mechanisms, particularly the Hexaglide mechanism, have received considerable attention from researchers due to their unique characteristics. The moving platform in such mechanisms is guided in parallel by several serial modules, resulting in high-speed and saccurate movement. Zhi-jiang and Ruo-chong have made significant contributions to the field by studying the kinematic model of a six-degree-of-freedom parallel mechanism, with a focus on the high-bending properties of its hinges\cite{du2012kinematics}. In 1997, Fang and Huang conducted a study on the kinematics of a mechanism that utilizes only three parallel actuators, as reported in their publication \enquote{Kinematics of a mechanism moving with only three parallel actuators}  \cite{fang1997kinematics}. Building upon this research, Patrick and Meng were able to further enhance the design of the mechanism in their 2002 study Kinematics of a mechanism moving with only three parallel actuators \enquote{Improvement of the design of a mechanism with three parallel actuators} \cite{patrick2002kinematics}. Additionally, Abtahi and Alasty examined the calibration of parallel robots through the implementation of motion constraints on the moving platform in their 2008 publication \enquote{Calibration of parallel robots through motion constraints on the moving platform} \cite{10.1115/IMECE2008-66163}. In their 2011 study, Xi and Fengfeng designed and analyzed a parallel mechanism with the capability to alter its configuration, as reported in their publication \enquote{Design and analysis of a parallel mechanism with the ability to change configuration} \cite{xi2011module}. Jakobovic and Domagoj, in their 2002 publication \enquote{Unsolved problems in direct and inverse kinematics of parallel mechanisms, with a special emphasis on the Stewart mechanism} \cite{jakobovic2002forward}, examined the unresolved issues in the direct and inverse kinematics of parallel mechanisms, particularly the Stewart mechanism. Other researchers have focused their attention on the Hexaglide mechanism. In their 2008 study \enquote{Kinematics and single point analysis of the Hexaglide parallel mechanism} \cite{abtahi2008kinematics}, Abtahi and Alasty investigated the kinematics and single point analysis of the Hexaglide parallel mechanism. Perng and Hsiao also performed a similar analysis for the inverse kinematics of the Hexaglide mechanism, as reported in their 1999 publication \enquote{Inverse kinematics analysis of the Hexaglide mechanism} \cite{perng1999inverse}. Furthermore, Honegger and Burdet were among the first researchers to apply adaptive control to the Hexaglide mechanism, as reported in their 1997 study \enquote{Adaptive control of the Hexaglide mechanism} \cite{honegger1997adaptive}. Building on the research of others, Alasti and Abtahi were able to successfully develop a CNC machine that utilizes the Hexaglide mechanism, as reported in their study \enquote{Development of a CNC machine with Hexaglide mechanism} \cite{10.1115/IMECE2008-66133}. Although these CNCs, which are based on parallel mechanisms, have advantages over traditional multi-axis perpendicular CNCs, they are not yet fully automatic. Therefore, the design and implementation of a controller that can create the necessary movements in the CNC is crucial. The G-Code generation software, which is commonly used in multi-axis CNC machines, cannot be used for CNC machines with parallel mechanisms due to their unique kinematics. Thus, the design and construction of a controller that can utilize standard G-Code with kinematic conversion for parallel CNC machines can be highly practical.

The structure of the remainder of this paper is as follows: In Section \ref{sec:Hexa_Robot}, we introduce the Hexaglide mechanism and its mathematical model. Section \ref{sec:Hexa_Kinematics} delves into the investigation of the geometrical parameters and the direct and inverse kinematics of the Hexaglide mechanism. Section \ref{sec:controller_design} provides a brief overview of the methods used by researchers to design and fabricate controllers for the Hexaglide mechanism. The results of the controller applied to the Hexaglide CNC are presented in Section \ref{sec:results}.\\

\section{Hexaglide Robot}\label{sec:Hexa_Robot}

\begin{figure}[!h]
\centering
\includegraphics[width=8.5cm]{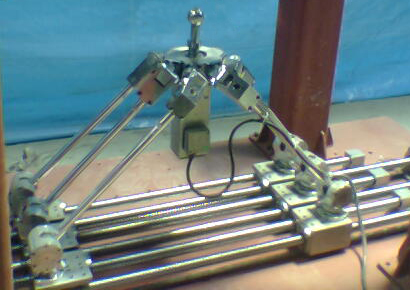}\\
\caption{Hexaglide Machine}
\label{fig:HexaReal}
\end{figure}

In this section, we present an overview of the Hexaglide mechanism, which is a crucial component in the development of the controller program algorithm. Figure \ref{fig:HexaReal} shows the manufactured Hexaglide machine.

\subsection {Introducing the components of the Hexaglide mechanism}\label{subsec:2-1}

A modular parallel mechanism is a type of mechanism that is composed of multiple individual modules, such as actuators, passive joints, rigid arms, a moving platform, and a tool center. The Hexaglide mechanism is a specific example of a generalized scaled parallel mechanism, which has a similar structure to the Stewart mechanism, but with the key difference that the length of the arms remains constant. The kinematic components of the Hexaglide mechanism include:
\begin{enumerate}
    \item Six arms with fixed lengths
    \item A moving platform
    \item Parallel rails
    \item Six prismatic joints on the rails
    \item Six spherical joints on the sliders (joints with three degrees of freedom)
    \item Six universal joints on the movable platform (joints with two degrees of freedom)
\end{enumerate}

\begin{figure}[!h]
\centering
\includegraphics[width=9cm]{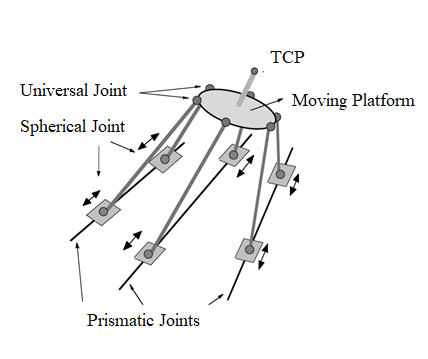}\\
\caption{Various components of the Hexaglide mechanism}
\label{fig:Hexa_Elements}
\end{figure}

In Figure \ref{fig:Hexa_Elements}, the various components of the Hexaglide mechanism are depicted. As shown, the arms connect the moving platform to six linear motors, which are situated on three linear rails through the use of universal joints. This arrangement allows for the precise movement of the platform and enables the mechanism to achieve a high level of accuracy and repeatability.

\subsection {The degree of freedom of the Hexaglide mechanism}\label{subsec:2-2}
In order to calculate the degree of freedom of the system, Gruber's relation has been used. In this relation, F is the degree of freedom of the studied mechanism and lambda is the degree of freedom of the space in which the mechanism works. L is the number of arms and j shows the number of joints in the mechanism. Also, the degree of freedom in the $i^{th}$ joint is $f_i$\cite{abtahi2008kinematics}:
y
To determine the degree of freedom of the system, we employ Gruber's relation. This relation states that the degree of freedom (F) of the mechanism being analyzed is equal to the degree of freedom lambda of the space in which the mechanism operates, multiplied by the number of arms (L) and subtracted by the number of joints (j) in the mechanism. Additionally, the degree of freedom in the $i^{th}$ joint is represented by $f_i$ \cite{abtahi2008kinematics}. 
 \begin{align}\label{eq:eq1}
    F = \alpha\times(I-j-1)+ \sum^{j}_{i=1} f_i
\end{align}

This relation is a widely used method for determining the degree of freedom of parallel mechanisms, and it was used in the research of Abtahi and Alasty for the kinematics of the Hexaglide mechanism.

In this mechanism, the Hexaglide mechanism, there are a total of 14 arms and 18 joints. Specifically, it consists of six joints with three degrees of freedom, six joints with two degrees of freedom, and six joints with one degree of freedom. With this information, we can use Gruber's relation to calculate the degree of freedom of the mechanism. By substituting the number of arms and joints, and the degree of freedom of each joint in the relation, we can obtain the degree of freedom of the Hexaglide mechanism.

\begin{equation*}
6\times(1-18-14) + 1\times6 + 2\times6 + 3\times6 = 6
\end{equation*}

As a result of applying Gruber's relation, it is determined that the degree of freedom of the Hexaglide mechanism is six. This indicates that the mechanism has the capability to move in all three dimensions of space and reach any point within its workspace. The Hexaglide mechanism is thus a versatile and highly capable parallel mechanism, making it well-suited for a wide range of applications.\\

\section{Hexaglide Kinematics}\label{sec:Hexa_Kinematics}

\subsection {Modeling and definition of initial parameters}\label{subsec:3-1}
\begin{figure}[!h]
\centering
\includegraphics[width=8cm]{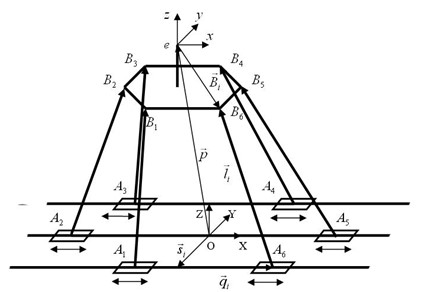}\\
\caption{Diagram illustrating the various components of the Hexaglide mechanism, including the six arms, parallel rails, linear motors, and universal joints connecting the moving platform}
\label{fig:Kinematics}

\end{figure}

As depicted in Figure \ref{fig:Kinematics}, in order to model the Hexaglide mechanism, two coordinate systems were utilized. This allows for a detailed representation of the mechanism's kinematics, enabling the analysis of its movement and behavior.

The first coordinate system used in this model is the general coordinate system, denoted as $O - XYZ$, with its origin located at the center of the middle rail. The $X$-axis of this coordinate system is aligned parallel to the direction of the rails, the $Y$-axis is perpendicular to the rails and parallel to the plane of the rails, and the $Z$-axis is perpendicular to the plane of the rails and points in the direction of the moving platform. The second coordinate system used is the local or mobile coordinate system, denoted as $e - xyz$, which is fixed on the moving platform. The orientation of its axes is depicted in Figure \ref{fig:Kinematics}. These two coordinate systems are used to accurately represent the kinematics of the Hexaglide mechanism.
As shown in the figure, the centers of the spherical and universal joints are represented by $A_i$ and $B_i$, respectively, where i ranges from 1 to 6. Unlike other parameters, the coordinates of the center of the universal joints, $B_i$, are defined relative to the local coordinate system. This allows for the specification of the vector connecting $A_i$ to $B_i$, and the length of the bases, $l_i$. The variables of linear joints are denoted by $q_i$, and the base position for spherical joints, represented by the middle point of the corresponding rail, is defined as $S_i$. The kinematic parameters of the Hexaglide mechanism are defined by the coordinates of the joints centers, the variables of linear joints, and the base position for spherical joints,  so we have:
 \begin{align}\label{eq:eq2}
    \overrightarrow{A_i} = \overrightarrow{S_i} + \overrightarrow{q_i}
\end{align}

where 
 \begin{align}\label{eq:eq3}
    &\overrightarrow{S_i} = [S_{ix}\quad S_{iy}\quad S_{iz}]^T \\ \nonumber
    &\overrightarrow{q_i} = [q_{i}\quad 0\quad 0]^T
\end{align}
On the other hand, the joint variable vector $q$ is defined as follows\cite{amiri2022impact}:
 \begin{align}\label{eq:eq4}
    \boldsymbol{q_i} = [q_{1}\quad q_{2}\quad q_{3}\quad q_{4}\quad q_{5}\quad q_{6}]^T
\end{align}

Furthermore, the position of the moving platform, or the origin of the local coordinate system, is represented by the vector $\overrightarrow{P}$, and the rotation angles of the moving platform around the $X$, $Y$, and $Z$ axes are denoted by $\alpha$, $\beta$, and $\gamma$, respectively. These kinematic parameters, along with the parameters mentioned earlier, are used to fully define the position and orientation of the Hexaglide mechanism in its workspace. The position vector of the moving platform is defined as follows:
 \begin{align}\label{eq:eq5}
    \boldsymbol{\chi} = [P_{x}\quad P_{y}\quad P_{z}\quad \alpha \quad \beta \quad \gamma]^T
\end{align}

To express the rotation of the moving platform with respect to the base coordinate system, a rotation matrix can be used. It is assumed that the moving platform first rotates around the $Z$-axis by an angle $\gamma$, followed by a rotation around the $Y$-axis by an angle $\beta$, and finally a rotation around the $X$-axis by an angle $\alpha$. The resulting rotation matrix can be defined as follows:

 \begin{align}\label{eq:eq6}
    \boldsymbol{R} = \boldsymbol{R_x (\alpha).R_y (\beta).R_z (\gamma)}
\end{align}

While $\boldsymbol{R_x (\alpha)}$, $\boldsymbol{R_y (\beta)}$, and $\boldsymbol{R_z (\gamma)}$  are the matrices of the initial period and are calculated in this way

 \begin{align}\label{eq:eq7}
\boldsymbol{R_x (\alpha)} =\begin{bmatrix}
1 & 0 & 0\\
0 & cos(\alpha) & -sin(\alpha)\\
0 & sin(\alpha) & cos(\alpha)
\end{bmatrix} \\ \nonumber
  \boldsymbol{R_x (\beta)} =\begin{bmatrix}
cos(\beta) & 0 & sin(\beta)\\
0 & 1 & 0\\
-sin(\beta) & 0 & cos(\beta)
\end{bmatrix} \\ \nonumber
  \boldsymbol{R_x (\gamma)} =\begin{bmatrix}
cos(\gamma) & -sin(\gamma) & 0\\
sin(\gamma) & cos(\gamma) & 0\\
0 & 0 & 1
\end{bmatrix} 
\end{align}\\

\subsection {Geometrical parameters of parallel robot of Hexaglide}\label{subsec:3-2} 

In the case of the Hexaglide parallel robot, it is assumed that the parallel rails are parallel to each other with high precision. As a result, the geometric parameters of the Hexaglide parallel robot include the coordinates of the spherical and universal joints, as well as the length of the bases. In total, there are 42 geometric parameters that define the Hexaglide parallel robot: 18 point components $S_i$, 18 point components $B_i$ relative to the local coordinate system, and 6 base lengths $l_i$. These parameters are necessary for the kinematic analysis and control of the Hexaglide parallel robot. Now the vector of geometric parameters can be defined as follows:
\begin{align}\label{eq:eq8}
    \boldsymbol{D} = [\overrightarrow{B}^T_1\quad \overrightarrow{B}^T_2\quad                       ...\quad \overrightarrow{B}^T_6\quad
    \overrightarrow{S}^T_1\quad \overrightarrow{S}^T_2\quad 
    ...\quad \overrightarrow{S}^T_6\quad l^T]^T
\end{align}
while
 
 \begin{align}\label{eq:eq9}
    &\overrightarrow{B_i} = [B_{ix}\quad B_{iy}\quad B_{iz}]^T \\ \nonumber
    &\overrightarrow{S_i} = [S_{ix}\quad S_{iy}\quad S_{iz}]^T \\ \nonumber
    & l = [l_1\quad l_2\quad l_3\quad l_4\quad l_5\quad l_6]^T \nonumber
\end{align}
In order to simplify and reduce the number of unknown parameters in the definition of coordinate devices, the following items have been included\cite{10.1115/IMECE2008-66133}:

\begin{enumerate}
    \item The origin of the general coordinate system is placed on the point $S_2$.
    \item The direction of the axis $X$ is parallel to the rails.
    \item The plane $X-Y$ passes through the point $S_3$.
    \item The origin of the local coordinate system is placed on the point $p$ or $e$.
    \item The plane $x-y$ is parallel to the plane $B_2-B_3-B_5$.
    \item The axis $x$ is parallel to the line $B_2-B_5$.
\end{enumerate}

According to the above:
 \begin{align}\label{eq:eq10}
    &S_{2x} = S_{2y} = S_{2z} = S_{3z} = 0  \nonumber \\
    &B_{2x} = B_{3z} = B_{5z}  \nonumber \\
    &B_{2y} = B_{5y}
\end{align}
 
Therefore, according to the three equations \ref{eq:eq10}, the number of independent parameters of the Hexaglide parallel robot, which has 42 geometric parameters, is reduced to 35.\\

\subsection {Inverse kinematics of parallel robot Hexaglide}\label{subsec:3-3} 
To evaluate the performance and simulate the behavior of the Hexaglide parallel robot, it is necessary to solve two problems: inverse kinematics and direct kinematics \cite{safavi2016force}. Inverse kinematics aims to determine the variable vector of the joints, $q$, for a given position vector of the moving platform, $\chi$. In general, the inverse kinematics of parallel robots has a unique solution and can be easily solved. Similarly, obtaining the inverse kinematic response of the Hexaglide parallel robot is also straightforward. On the other hand, direct kinematics is used to calculate the position and orientation of the end-effector (moving platform) for a given set of joint variables. The following vector equation can be written for each base of the parallel Hexaglide robot:
 \begin{align}\label{eq:eq11}
    \overrightarrow{l_i} = \overrightarrow{P} + R\overrightarrow{B_i} - \overrightarrow{q_i} - \overrightarrow{S_i}
\end{align}
While $\overrightarrow{P}$ and $R$ are defined from $\chi$ and $\overrightarrow{B_i}$, $\overrightarrow{S_i}$ and $l_i$ are among the geometric parameters. Using the above equation:
 \begin{align}\label{eq:eq12}
    l^2_i = \overrightarrow{l^T_i}\overrightarrow{l_i} = (\overrightarrow{P} + R\overrightarrow{B_i} - \overrightarrow{q_i} - \overrightarrow{S_i})^T(\overrightarrow{P} + R\overrightarrow{B_i} - \overrightarrow{q_i} - \overrightarrow{S_i})
\end{align}
If the vector $\overrightarrow{b_i}$ is defined as follows:
\begin{align}\label{eq:eq13}
    \overrightarrow{b_i} = \overrightarrow{P} + R\overrightarrow{B_i} - \overrightarrow{q_i} - \overrightarrow{S_i}
\end{align}
According to the second part of equation \ref{eq:eq3} and by extracting the value from equation \ref{eq:eq12}, the following equation is deduced:
\begin{align}\label{eq:eq14}
    q_i = IK(\chi, D) = b_{ix} + h_i\sqrt{l^2_i - b_{iy}^2 - b_{iz}^2}
\end{align}

The values $h_i$ can take two values $1$ and $-1$. Now, to select the indicated stable answer, the values $h_i$ are selected as follows:
\begin{equation*}
h_i = \Bigg\{ \begin{matrix}
           1 & for & i=4,5,6\\
           -1 & for & i=1,2,3
\end{matrix}
\end{equation*}

Equation \ref{eq:eq14} can be used to find the variable vector of the joints according to the position vector of the moving platform.\\

\subsection {Direct kinematics of Hexaglide parallel robot}\label{subsec:3-4}

The goal of direct kinematics is to calculate the position vector of the moving platform $\chi$ for a given joint variable vector $q$. In contrast to inverse kinematics, the direct kinematics of parallel robots does not have a unique solution and finding the solution through analytical methods can be highly challenging, if not impossible in many cases. There are non-numerical approaches to solve direct kinematics for robots with less than 4 degrees of freedom in certain special cases and with simplifying assumptions. However, for robots with high degrees of freedom, such as the Hexaglide parallel robot with 6 degrees of freedom, numerical methods are the most efficient and effective approach. One popular method is the Newton-Raphson method. This method involves obtaining the Jacobian matrix of the Hexaglide parallel robot, using inverse kinematics and an initial guess of the direct kinematic response, and then finding the final solution numerically.

The Newton-Raphson method can be summarized in the following steps \cite{MerletForthcoming-MERLRP-2}:
\begin{enumerate}
\item Consider an initial state vector $\chi$.
\item $q_i = IK(\chi, D)$ to be calculated
\item If $\|q^d - q\| > \epsilon$, then the new value $\chi$ is calculated using \ref{eq:eq15} and refer to step 2, otherwise the $\chi$ is the answer of direct kinematic.
\end{enumerate}
\begin{align}\label{eq:eq15}
    \chi = \chi + J(q^d - q)
\end{align}
The variable vector of the optimal input joints, represented by $q^d$, is an essential component in robot control. It is used in conjunction with the permissible error, denoted by $\epsilon$, and the Jacobian matrix, represented by $J$, to ensure precise and accurate calculations. The Jacobian matrix, which represents the relationship between the robot's input joints and its end-effector velocities, is obtained by analyzing the speed of the robot. It plays a crucial role in determining the optimal input joints and ensuring that the robot's movements are in line with the desired performance specifications. In summary, the combination of $q^d$, $\epsilon$, and $J$ is fundamental in achieving precise and accurate robot control.\\

\section{Designing and manufacturing of the controller}\label{sec:controller_design}

\subsection {Definition of G-Code}\label{subsec:4-1}
G-Code is a type of programming language that is used to control CNC machines. It is a text file that contains the necessary commands for a specific path or operation on the CNC machine. There are various software programs available for creating G-Code, such as CATIA, NX, and SolidWorks, which can be used to model the desired object and generate the necessary G-Code for the CNC machine. Furthermore, there are specialized software programs specifically designed for G-Code generation, such as Mastercam, which can take a 3D model in a standard format and, based on specific conditions, calculate the path of the tool and generate the corresponding G-Code for the CNC machine. These software programs perform various calculations related to tool movement speed, surface smoothness, tool diameter, and the shortest path and loading times for a surface.

G-Code commands are generally categorized into five groups:
\begin{enumerate}
    \item Motion commands related to machining, which typically begin with the letter G.
    \item Coordinate and angle-related commands, which typically begin with the letters X, Y, Z, A, B, I, J, and K.
    \item Non-moving commands related to machining, which typically begin with the letter G, such as commands for compensating for tool length and diameter, changing the coordinate plane, etc.
    \item Miscellaneous commands, which typically begin with the letter M and include commands for other functions such as spindle speed (S) and tool feed rate (F).
    \item Other commands that do not fit into the above categories, but still have a specific function in controlling the CNC machine.
\end{enumerate}
This categorization of G-Code commands helps in understanding their functionality and how they are used in controlling the CNC machine.

An important consideration when choosing a software for generating G-Code is its capability to support five-axis machining independently. This is important because the Hexaglide mechanism CNC machine, due to its kinematic coupling, cannot use standard G-Code. Therefore, the ability to convert standard G-Code to G-Code specifically tailored for the Hexaglide mechanism can be highly beneficial.\\

\subsection {Controller Design}\label{subsec:4-2}
\begin{figure}[!h]
\centering
\includegraphics[width=8cm]{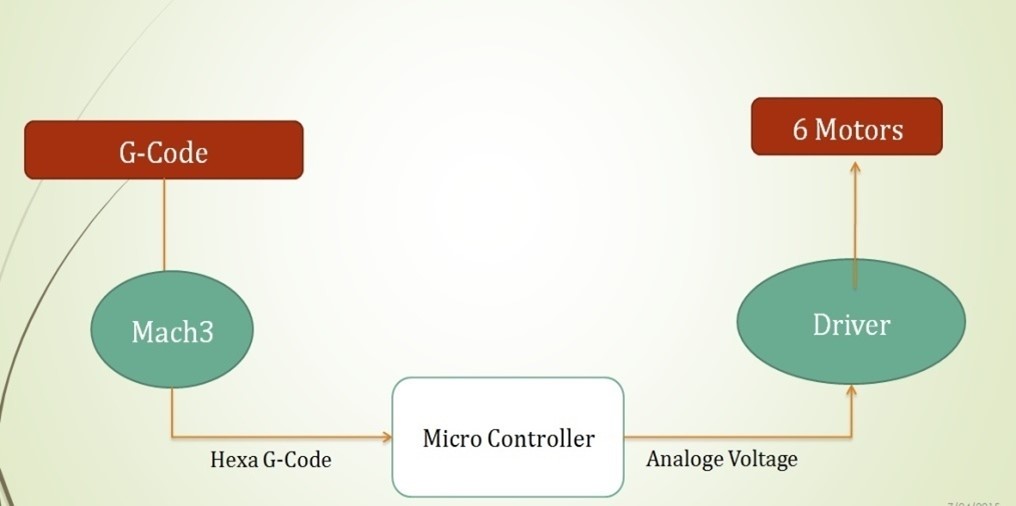}\\
\caption{Schematic of the research path}
\label{fig:Schematic}

\end{figure}

After generating the G-Code using the code software, it is inputted into the Mach3 software. Mach3 is a controller simulator that is commonly used in regular CNCs. It takes the G-Code as input and generates the pulses required for the motor drivers through the printer port. With the removal of printer ports from modern computers, necessary modifications have been made to the Mach3 software such that it now outputs the pulse creation commands via the USB port. The USB port is connected to a microcontroller which then analyzes the commands and generates the necessary pulses for the drivers. The Mach3 software also has the capability to be modified through the use of plugins, allowing the user to customize the input and output of the software according to their needs. By modifying the instructions for movement, which is considered as the output of the Mach3 program, to align with the inverse kinematics of the Hexaglide mechanism, it is possible to convert standard G-Code to G-Code specific to the Hexaglide mechanism by utilizing the Mach3 software. The overall path of the research is illustrated in Figure \ref{fig:Schematic}.

In summary, the process of designing the controller for the Hexaglide CNC machine involves the following steps: First, G-Code is generated using CAD software. Then, it is inputted into the Mach3 program for analysis. By writing the corresponding code, the movement commands are converted according to the inverse kinematics of the Hexaglide mechanism, resulting in the necessary movement commands for the Hexaglide mechanism. These commands are then sent to the microcontroller through the USB port, where they are analyzed and used to generate the pulses required to control the motors. It is important to note that the drivers in the Hexaglide CNC machine work with analog signals, so a digital to analog conversion is necessary. Once the drivers receive the desired input, they are able to start the motors connected to the Hexaglide mechanism.\\

\subsection {Creating the electrical board}\label{subsec:4-3}
\begin{figure}[!h]
\centering
\includegraphics[width=8cm]{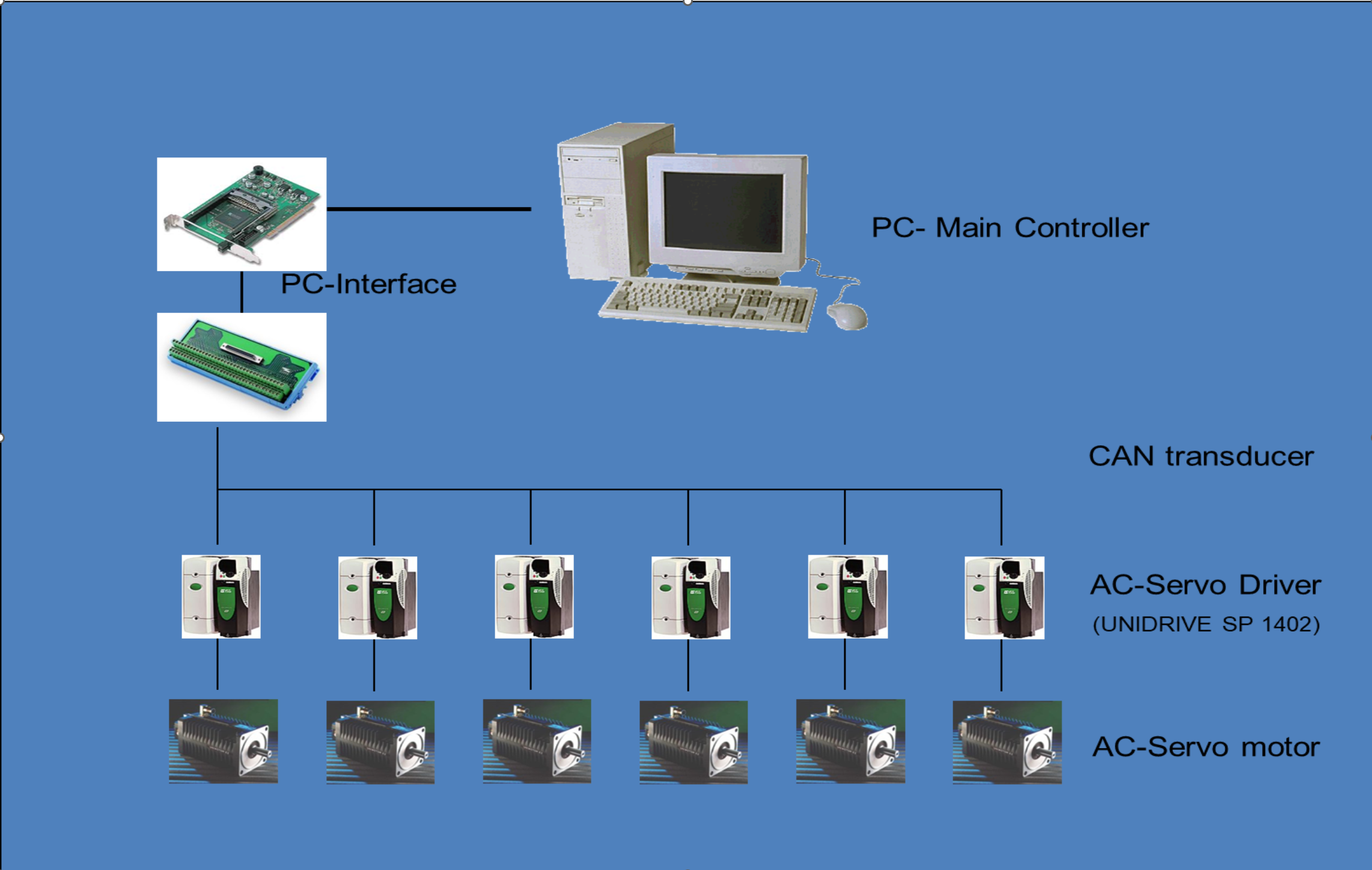}\\
\caption{Placement of the controller in the circuit of the CNC Hexaglide device}
\label{fig:Circle}

\end{figure}

As illustrated in Figure \ref{fig:Circle}, the electrical board plays a crucial role in interpreting the commands generated by the Mach3 program and converting them into the necessary commands to control the movement of the motors in the Hexaglide mechanism.

\begin{figure}[!h]
\centering
\includegraphics[width=8cm]{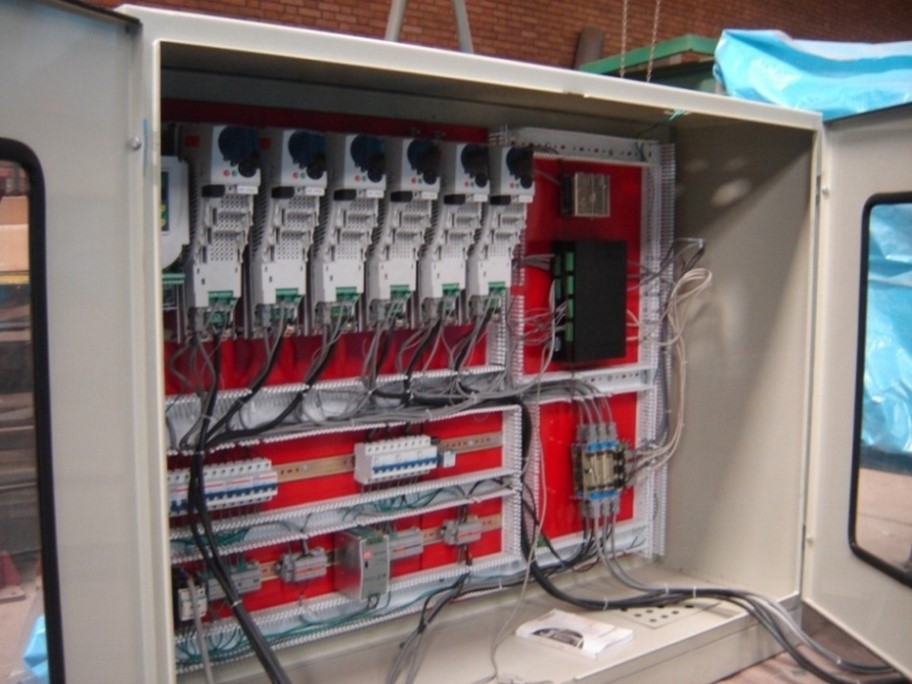}\\
\caption{Components of the controller circuit of the CNC Hexaglide device}
\label{fig:RealController}
\end{figure}

The heart of this board's processor is the microcontroller, which processes the movement commands for the mechanism received from the computer via the serial port. The motor drivers regulate the motor's speed based on the voltage difference applied to them. The microcontroller is responsible for controlling the position of the mechanism by generating the necessary pulse for the motors. It does this by utilizing internal digital-to-analog converters to create the appropriate voltage levels for the drivers. However, since the maximum voltage generated by the microcontroller is 3.3 volts, while the drivers require a voltage difference of 10 volts to operate, an op-amp is used to amplify the voltage level from the microcontroller. As seen in Figure \ref{fig:RealController}, the controller board is located next to the motor drivers.
\cite{huang2017active} \\
 
\section{Results}\label{sec:results}
The CNC Hexaglide equipped with the designed controller has successfully achieved the desired movement patterns. The system has been tested and validated through various experiments. In one experiment, the system was tested for straight-line movements with different step sizes. The results showed that the system was able to accurately follow the input commands and produce the desired movement patterns with negligible errors.

Furthermore, the controller was tested for circular and curved trajectories. The system was able to accurately follow the input commands and produce smooth curves with minimal deviations from the desired path. The accuracy and precision of the system were further improved by tuning the controller parameters.

In addition to the above experiments, the system was also tested for complex trajectories and shapes, such as ellipses and polygons. The results demonstrated that the controller was capable of generating accurate and smooth trajectories for a wide range of shapes.

The controller also offers several advanced features, such as real-time monitoring and control of the CNC machine, and the ability to adjust the system parameters on-the-fly. These features provide enhanced flexibility and control to the user.

Overall, the designed controller for CNC Hexaglide has proven to be highly effective in generating precise and accurate movements, while also providing advanced features and flexibility to the user. The results of the experiments demonstrate the potential of this controller for use in a wide range of CNC applications.
\begin{figure}[!h]
\centering
\includegraphics[width=8cm]{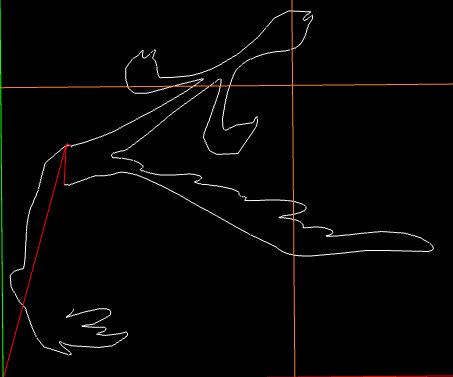}\\
\caption{Generated pattern by CNC Hexaglide}
\label{fig:genPattern}
\end{figure} 

\section{Conclusion}\label{sec:conclusion}

In conclusion, the design and control of a parallel mechanism such as the Hexaglide requires a thorough understanding of its kinematics, as well as the use of appropriate mathematical tools and software. The Hexaglide mechanism, with its six arms and fixed lengths, offers six degrees of freedom and is capable of covering the entire three-dimensional space. The inverse and direct kinematics of the Hexaglide mechanism were investigated, and the Newton-Raphson method was used to solve the direct kinematics problem. The G-Code was generated using CAD software, and the Mach3 program was used to convert the G-Code into the necessary movement commands for the Hexaglide mechanism. These commands were then processed by the microcontroller on the electrical board, which converted them into the necessary pulses for the motor drivers. The electrical board also included an op-amp to amplify the voltage levels from the microcontroller to the required levels for the motor drivers. Overall, the Hexaglide mechanism has proven to be a versatile and effective parallel mechanism, and with the use of proper design and control methods, it can be utilized in various applications such as CNC machines.

%---------------------------------------------------------------------------------------------------------
\balance
\bibliographystyle{IEEEtran}
\bibliography{ref}{}

%---------------------------------------------------------------------------------------------------------
%---------------------------------------------------------------------------------------------------------
\end{document}